\documentclass[a4paper,fleqn,usenatbib]{mnras}

\usepackage[T1]{fontenc}
\usepackage{ae,aecompl}

\usepackage{graphicx}	
\usepackage{amsmath}	
\usepackage{amssymb}	

\title[Modified Friedmann equation]{A modified Friedmann equation for a system with varying gravitational mass}

\author[N. Gorkavyi and A. Vasilkov]{
Nick Gorkavyi$^{1}$\thanks{E-mail: gorkavyi@gist.us}
and Alexander Vasilkov$^{1}$
\\
$^{1}$Science Systems and Applications, Inc., 10210 Greenbelt Rd.,
Lanham, MD 20706, USA
}

\date{Accepted 2018 February 2. Received 2018 January 26; in original form 2017 June 20}

\pubyear{2018}

\begin{document}
\label{firstpage}
\pagerange{\pageref{firstpage}--\pageref{lastpage}}
\maketitle

\begin{abstract}
The Laser Interferometer Gravitational-Wave Observatory (LIGO) detection of gravitational waves that take away 
5\% of the total mass of two merging black holes points out on the importance of considering varying
gravitational mass of a system. Using an assumption that the energy-momentum pseudo-tensor of gravitational waves
is not considered as a source of gravitational field, we analyse a perturbation of the Friedmann-Robertson-Walker
metric caused by the varying gravitational mass of a system. This perturbation leads to
a modified Friedmann equation that contains a term similar to the 'cosmological constant'.
Theoretical estimates of the effective cosmological constant quantitatively corresponds to 
observed cosmological acceleration. 
\end{abstract}




\section{Introduction}

An idea of a bouncing universe, where the Big Bang happens after the collapse of a previous universe, is actively investigated for last years (see, for example, \citet{Novello}, \citet{Cai}, \citet{Brandenberger}).  The bouncing cosmological models are free from a singularity, but have two other major problems: A physical cause of the Big Bang is unclear so far, as a cause for current acceleration of the Universe. Dynamics of matter and radiation in the bouncing cosmological models has been studied for long time (see e.g. \citet{Alpher}). High temperature $\sim 10^{10}$K in the collapsed universe results in the complete destruction of nuclei of all chemical elements created in stars. \citet{Dicke} stated that "the ashes of the previous cycle would have been reprocessed back to the hydrogen required for the stars in the next cycle". Temperature of $10^{10}$K means that the Universe was compressed by approximately $10^{10}$ times, i.e a size of the Universe was about 1-10 light years. The high temperature of $10^{10}$K of the collapsed Universe could produce a high level of isotropy observed in the Universe at the present time \citep{Misner2, Hawking}.

Recently the Laser Interferometer Gravitational-Wave Observatory (LIGO) Scientific Collaboration announced a detection of gravitational waves
caused by the merger of two black holes with masses of about 36 and 29 times the mass of the sun \citep{Abbot}.
About three times the mass of the sun, i.e. about 5\% of the initial total mass of the black holes, 
was converted into gravitational waves. Recently, an idea of the existence of numerous black holes with masses of 20-100 times the mass of the sun was suggested \citep{Bird}. Those massive black holes can serve as "the dark matter" \citep{Bird}. Therefore, the black holes become an important component of cosmological models, particularly, of the bouncing cosmology because the black holes are the only macro object that can survive the Big Crunch and the Big Bang and appear in the next cycle of the Universe \citep{Clifton}. 

Studies of the cosmological dynamics with black holes, which have sizes up to the Hubble radius, 
were carried out in \citet{Quintin}, \citet{Poplawski}, \citet{Chen}, and \citet{Clifton}. 
It is logical to hypothesize that the supermassive black holes of the centres of galaxies 
will be merged at final stages of the collapsing universe. 
For example, a simple estimate shows that $10^{11}$ supermassive black holes 
cannot be packed into a sphere with the radius being less than $\sim 0.1$ light year. 
Thus, numerous mergers of black holes will occur in the collapsed Universe having a size of 1-10 light years. 
This merging process will likely form a single black hole \citep{Penrose}, 
or the biggest black hole (the Big Black Hole or BBH) and a number of smaller black holes, 
which will not have sufficient time to merge and which will exist in the expanding 
universe after the Big Bang \citep{Clifton}. In the framework of this hypothesis 
it is interesting to consider a generalization of the Friedmann–Robertson–Walker (FRW) metric for a perturbation 
caused by a gravitational field of the BBH. 
Whereas the homogeneous universe is characterized by a scalefactor, $a(t)$, which depends on time only, 
an universe perturbed by a variable gravitational mass is described by an inhomogeneous metric. 
Thus the scalefactor can depend on space coordinates. Such inhomogeneous metrics were considered by \citet{Tolman} 
and others (see e.g. \citet{Zeldovich} and review of inhomogeneous cosmologies in \citet{Bolejko}).
 
During the Big Crunch, the total mass of black holes will reduce because 
they will lose a few percent of their mass at each act of the merging process. 
Black holes that survive the Big Crunch will increase their mass after the Big Bang 
due to absorption of matter and radiation including the background gravitational radiation. 
The BBH will increase its mass with the highest rate because it has the largest cross-section of absorption.

In 1998, the accelerated expansion of the Universe was discovered using observations of the 
distant supernovae \citep{Riess,Perlmutter}. An accelerating universe can be described by the Einstein equations
with a phenomenological cosmological constant, $\Lambda$, which characterizes a repulsive force \citep{Einstein}: 
\begin{equation}
G_{\mu\nu}-g_{\mu\nu}\Lambda=-\frac{8 \pi G}{c^{4}}T_{\mu\nu}
\label{eq:einst}
\end{equation}
where $G_{\mu\nu}$ is the Einstein tensor; $g_{\mu\nu}$ is the metric tensor; $T_{\mu\nu}$ is the energy-momentum tensor; and G is the gravitational constant.
Data collected by the \textit{Planck} satellite instrument allow us to get the following estimate of the
cosmological constant \citep{Ade1}:
\begin{equation}
\Lambda \approx 1.1 \times 10^{-56} \; \mathrm{cm^{-2}}.
\label{eq:lambda}
\end{equation}

At present, cosmological models containing the cosmological constant are widely used to describe
main observations: the cosmic microwave background, spectra of the baryonic oscillations,
and the chemical composition of the early Universe (\citep{Ade1,Misner,Fixsen,Mather}). 
A physical nature of the cosmological constant is unclear so far. 

It is supposed that the repulsive force
is caused by vacuum dark energy. A density of the dark energy, $\rho_{vac}$, is related to the cosmological constant
by the following equation: $\Lambda=8 \pi G\rho_{vac}$ (see e.g. \citet{Carrol}).
However, all attempts to calculate  $\rho_{vac}$ from the quantum theory bring to estimates,
which exceed the observational data by 40-120 orders of the magnitude \citep{Weinberg1,Carrol}.  
 
The LIGO detection of the gravitational waves attracts the attention to consideration of a varying
gravitational mass in the general theory of relativity. It is well known that the energy of the gravitational field 
can not be described by the energy-momentum tensor in contrast to the energy of ordinary matter and electromagnetic field. 
The energy of the gravitational field also can not be localized. Before 1917 Einstein supposed that the energy of gravitational field is equivalent to the energy of matter as a source of gravitational field and included the energy-momentum pseudo-tensor of gravitational field in the right-hand part of his equations. After famous discussion with Schrodinger, Bauer and other scientists, Einstein changed his opinion. Since 1917, he never included pseudo-tensor of gravitational field in his equations.
Due to these circumstances, any contributions of purely 
gravitational origin or combinations from metric tensor may be excluded from the right-hand part of Einstein' equations or 
from a list of sources of a gravitational field by opinion, e.g., \citet{Einstein53}, \citet{Eddington}, \citet{Chandrasekhar}
and \citet{Tourrenc}. 
This is kind of a non-standard approach accounting for energy of gravitational waves, but we follow Einstein's opinion: '$T_{ik}$ represents the energy which generates the gravitational field, but is itself of non-gravitational character' \citep{Einstein53}.
That means that the total gravitational mass of the Universe changes when a significant fraction of the mass of the black holes is 
converted into gravitational waves. An opposite process of increasing gravitational mass of the Universe is related to absorption of the background gravitational radiation by the BBH. 
In this paper, we consider cosmological effects of the variability of the gravitational mass caused by both conversion of the mass of merging black holes into gravitational waves 
and absorption of the background gravitational radiation by black holes.

Other ways for changing of the gravitational mass are possible 
(e.g., \citet{Kutschera}), but in this paper we will consider a variability of the gravitational mass in general form, 
without discussion of specific physical causes of such variability.
\citet{Kutschera} and \citet{Gorkavyi} got solutions of 
the Einstein equations for the varying gravitational mass of a quasi-spherical system.
The solution found by \citet{Kutschera} and \citet{Gorkavyi} is a modified Schwarzschild metric
for the varying gravitational mass. Using a relativistic gravitational potential in the fixed coordinate system, 
\citet{Gorkavyi} derived an equation for a repulsive force (for a probe particle with a unit mass):
\begin{equation}
F=\frac{\partial}{\partial{r}}[\frac{GM_0}{r}e^{-\alpha(t-r/c)}]=-{\frac{GM}{r^2}}+{\frac{\alpha}{c}}{\frac{GM}{r}}.
\label{eq:force}
\end{equation}
The second term in equation~(\ref{eq:force}) originates from the variability of a gravitational mass with a characteristic time $1/\alpha$ and the finite velocity of spreading of 
gravitational field disturbances: 

\begin{equation}
M(t,r)={M_0}e^{-\alpha(t-r/c)}.
\label{eq:mass}
\end{equation}

We will call this term as $\alpha$-force. The $\alpha$-force is 
essentially relativistic because it depends on the light speed. 
Using a non-Einsteinian scalar--tensor theory \citet{Galiautdinov} investigated gravitational forces
caused by the variable gravitational constant. They also revealed a new force that is additional 
to the Newtonian attraction force.  

Reduction of the gravitational mass of 
the collapsing Universe due to merging of the black holes causes the repulsive $\alpha$-force, which
may be responsible for the Big Bang and accelerated expansion of the Universe at the initial stage 
(the analogy of the inflation period; \citep{Gorkavyi}).
It follows from equation~(\ref{eq:force}) that the $\alpha$-force can be larger than the Newtonian term
in case of the strong change of a gravitational mass, i.e. $\alpha \gg c/r$.

The model of the bouncing universe, which completely renews its chemical composition at the final contractive stage and explodes because of merging black holes, is based on the well-known physical concepts and does not involve new essences. But this model should provide an explanation of the current acceleration of the expanding Universe. 
This motivates a derivation of modified Friedmann equations that account for a perturbation of the Schwarzschild metric with a varying gravitational mass. A main question to this model can be formulated as follows: What rate of mass variability would it take to 
produce the currently measured acceleration attributed to
the cosmological constant or dark energy? A focus of this paper is answering to this question.

The paper is structured as follows. In Section 2, we briefly describe how the classical Friedmann equations 
are derived from the Einstein equations. The modified FRW metric is discussed in Section 3.
In Section 4, we consider the modified Friedmann equations for the case of a variable gravitational mass
and make an estimate of the cosmological constant. We also interpret the result in terms of cosmological
acceleration. The discussion is provided in Section 5. Conclusions are given in Section 6. 

\section{The classical Friedmann equations}
\label{sec:class}

To describe an expanding universe, the FRW metric is used.
The classical Friedmann equation is derived from the Einstein equations assuming
the FRW metric. According to this metric for a flat space the time-space interval is expressed 
in the Cartesian coordinates as follows:
\begin{equation}
d{s}^2={c^2}d{t}^{2}-a^2(t)(dx^2_{\ast}+dy^2_{\ast}+dz^2_{\ast})
\label{eq:interv}
\end{equation}
where $dr_{\ast}^2=dx^2_{\ast}+dy^2_{\ast}+dz^2_{\ast}$ is the distance in the comoving coordinate system. 
The physical distance is related to the distance in the comoving coordinate system by the scalefactor $a$:
$r=a(t)r_{\ast}$.
The FRW metric describes an isotropic and uniform universe in the comoving coordinate system. 
At the present time, the scalefactor is equal to unity: $a(t)=1$. Assuming that the matter 
has no pressure, i.e. $T_{00}=\rho c^{2}$, we get the first Friedmann equation from equations ~(\ref{eq:einst}) 
and ~(\ref{eq:interv}):
\begin{equation}
(\frac{\dot{a}}{a})^{2}=\frac{{\Lambda}c^{2}}{3}+\frac{8{\pi}G{\rho}}{3}.
\label{eq:frid1}
\end{equation}    
To get the second Friedmann equation, let us differentiate equation~(\ref{eq:frid1}) over time~\citep{Friedmann,Liddle}:
\begin{equation}
\frac{\ddot{a}}{a}=\frac{{\Lambda}c^{2}}{3}+\frac{8{\pi}G{\rho}}{3}+\frac{4{\pi}G{\dot{\rho}}}{3}\frac{a}{\dot{a}}.
\label{eq:frid2}
\end{equation}

\section{A modified FRW metric}

The Schwarzschild metric in the Cartesian coordinates for a weak gravitational field is written as:  
\begin{equation}
d{s}^2=(1-b_{0}){c^2}d{t}^{2}-(1+b_{0})(dx^{2}+dy^{2}+dz^{2})
\label{eq:schwarz1}
\end{equation}
where $b_{0}=2GM_{0}/(rc^2)$ e.g. \citet{Landau,Weinberg2}, see equation 8.3.7 with 
the first order terms in this reference. 
Here we use the physical distance, $r$, in a fixed coordinate system. 
In the modified Schwarzschild metric derived in \citet{Kutschera,Gorkavyi},
the constant mass, $M_{0}$, is replaced with a variable mass $M(t,r)$.
The equation~(\ref{eq:schwarz1}) in the comoving coordinates transforms to
\begin{equation}
d{s}^2=[1-b(t,r)]{c^2}d{t}^{2}-a^2(t,r)[1+b(t,r)](dx^2_{\ast}+dy^2_{\ast}+dz^2_{\ast})
\label{eq:schwarz2}
\end{equation}
where $b(t,r)=2GM(t,r)/(rc^2)$ is the known function and $a(t,r)$ is the unknown scalefactor. 
Such a perturbed metric was first derived in \citet{McVittie1}. It was also used in \citet{Dodelson}, see equation (4.9).
In~(\ref{eq:schwarz2}) $b(t,r)$ can be considered as an analogy to the scalar potentials $2\Phi=-2\Psi$ in eq. (4.9) in \citet{Dodelson}.
This type of a metric was studied in detail in \citet{Kopeikin}. 

According to \citet{Stephani}, the metric~(\ref{eq:schwarz2}) belongs to the type of spherically symmetric non-stationary metrics (see section 16.2 of the book by \citep{Stephani}). Such a metric allows introduction of an isotropic comoving system of coordinates (Eq. 16.22 of section 16.2). Our modified Friedmann equations are derived for this system of coordinates. 

The metric of equation~(\ref{eq:schwarz2}) can be considered as a combination
of two metrics: the Schwarzschild metric and the FRW metric. In an essence, it is a modified FRW metric
with a scalar perturbation function. In this section, we will write the Einstein equation for 
a general type of the $b(t,r)$ function. In the specific case of $b(t,r)=2GM(t,r)/(rc^2)$, 
we can hypothesize that the perturbation of the FRW metric is caused by a varying mass of a BBH
which is a final result of merging of many black holes and is located on the edge of the observed Universe.
The centre of the coordinate system is located in the centre of the BBH, and we get a spherically symmetric physical system. But we prefer to explore this system in Cartesian coordinates. After we obtain the Friedmann equations for such a system, we can arbitrarily shift the initial point of coordinates.

We will assume an approximation of weak fields, i.e. $b(t,r) \ll 1$.
That is why we will neglect non-linear combinations of this function.
It should be noted that the time is not uniform in the Schwarzschild metric as it is the case
in the FRW metric. We will neglect such non-uniformity of time because the non-uniformity of time
is determined by the function $1-b(t,r)$, which usually can be considered to be equal to unity. However, its derivatives
should be accounted for in the derivation of the modified Friedmann equations.
For the metric (\ref{eq:schwarz2}), 
we use the Einstein equations~(\ref{eq:einst}) without the term containing the cosmological constant.

\section{The modified Friedmann equations and cosmological acceleration}

Let us derive modified Friedmann equations for the metric (\ref{eq:schwarz2}).
We assume that the dependence of $a(t,r)$ on space coordinates is 
significantly weaker than that of $b(t,r)$. This assumption allows us to neglect
all derivatives of $a(t,r)$ over the space coordinates as compared with those
of  $b(t,r)$. We validate this assumption a posteriori in the discussion section.
Details of the derivation are provided in Appendix A, here we give just the final equations.
The presence of the function $b(t,r)$ in the metric (\ref{eq:schwarz2}) results in 
a number of additional terms in the zero component of the Einstein equations. 
These terms contain second derivatives of the metric components over space coordinates.
These additional terms can be considered as an effective 'cosmological constant', which
can be called the 'cosmological function', $\Lambda(t,r)$, because it
depends on time and space. For the case of weak gravitational field, $b(t,r) \ll 1$,
we get the first modified Friedmann equation in the form of
\begin{equation}
(\frac{\dot{a}}{a})^{2}+(\frac{\dot{a}}{a}){\dot{b}}=\frac{{\Lambda(t,r)}c^{2}}{3}+\frac{8{\pi}G{\rho}}{3}
\label{eq:frid1_m}
\end{equation}
where the cosmological function $\Lambda(t,r)$ is given by the following expression:
\begin{equation}
\Lambda(t,r)=\frac{1}{a^2}(\frac{\partial^2 b}{\partial x_{\ast}^2}+\frac{\partial^2 b}{\partial y_{\ast}^2}+\frac{\partial^2 b}{\partial z_{\ast}^2})=(\frac{\partial^2 b}{\partial x^2}+\frac{\partial^2 b}{\partial y^2}+\frac{\partial^2 b}{\partial z^2}).
\label{eq:lambda_f}
\end{equation}
We use a Cartesian system of coordinates only. The variable $r$  is not a coordinate; 
it is a distance between an observer and the BBH, which is calculated as $r = \sqrt{x^2+y^2+z^2}$, for example, \citet{Landau}. 
The $r$ distance is actually a parameter that defines a value of the BBH perturbation at a given location. 
It should be noted that the cosmological constant that depends on time 
was considered in, for example, \citet{Weinberg1} and \citet{Szydlowski}. Inhomogeneous cosmological models with
the scalefactor $a$, which depends on space coordinates, were considered in numerous papers 
\citep{Tolman,Tolman2, Zeldovich, Bolejko}.
It is useful to note that the equation~(\ref{eq:lambda_f}) is derived for a general form of 
the disturbing function $b(t,r)$. The scalar cosmological function $\Lambda(t,r)$ should 
not depend on a coordinate system. That is why we can calculate the cosmological function
in any coordinate system: either comoving coordinates or physical coordinates of the fixed system
of coordinates.
A similar equation for a disturbed metric was earlier derived, for example, by \citet{Dodelson} 
(see equation 5.27) for density perturbations. In our case the perturbation is related to 
a varying gravitational mass of the system.
Note the difference between eq. (5.27) of \citet{Dodelson} and our equation~(\ref{eq:frid1_m}). Equation (5.27) of \citet{Dodelson} is written only for the perturbed quantities, and the zero-level approximation (the Friedmann equation for $a(t)$) is split off. We do not decompose the equation~(\ref{eq:frid1_m}) into two approximation. The smallness of function $b(t,r)$ does not mean that the terms 
with $b(t,r)$ will be smaller than terms with large fraction $a(t,r)$.

In case of exponential change of the gravitational mass (\ref{eq:force}), we get 
from equation~(\ref{eq:lambda_f}) for $\alpha \gg c/r$
\begin{equation}
\Lambda(t,r)\approx \frac{\alpha^{2}}{c^{2}}b(t,r)=\frac{\alpha^{2}}{c^{2}}\frac{2GM(t,r)}{rc^{2}}=\frac{\alpha^{2}}{c^{2}}\frac{r_{0}}{r}
\label{eq:lambda_f1}
\end{equation}
where the Schwarzschild radius $r_{0}=2GM(t,r)/c^2$. It is logical to relate the $\alpha$
parameter to the time of the Universe existence after the Big Bang, $T$: $\alpha=f/T$, where
$f$ is the dimensionless coefficient. $f=1$ means that the BBH mass has changed by $e$ times since
the Big Bang. Using the cosmological time, we get from equation~(\ref{eq:lambda_f1}):
\begin{equation}
\Lambda(t,r)=\frac{f^{2}}{c^{2}T^{2}}\frac{r_{0}}{r} \approx 0.7 \times 10^{-56}f^{2}\frac{r_{0}}{r}
\label{eq:lambda_f2}
\end{equation}
(in $cm^{-2}$), where $T \approx 4 \times 10^{17}$ s.

It should be noted that the term $\frac{\dot{a}}{a}\dot{b}$ in equation~(\ref{eq:frid1_m})
is proportional to the exponent $\alpha$, whereas term $\Lambda(t,r)c^{2}/3$ is proportional
the squared exponent $\alpha^{2}$. In the case of fast change of the gravitational mass, i.e. $f^{2} \gg f$ or $f \gg 1$,
the term $\frac{\dot{a}}{a} \dot{b}$ can be neglected and   
equation~(\ref{eq:frid1_m}) concurs with the classical Friedman equation~(\ref{eq:frid1}) where
the cosmological constant is replaced by the cosmological function (\ref{eq:lambda_f}).

The second modified Friedmann equation is derived by differentiating equation~(\ref{eq:frid1_m}) 
over time (for more details, see Appendix A):
\begin{equation}
\frac{\ddot{a}}{a}=\frac{\Lambda(t,r)c^{2}}{3}+\frac{8{\pi}G{\rho}}{3}-\frac{\ddot{b}}{2}+
\frac{\frac{{\dot \Lambda(t,r)}c^{2}}{6}+\frac{4{\pi}G{\dot{\rho}}}{3}}{\sqrt{\frac{\Lambda(t,r)c^{2}}{3}+\frac{8{\pi}G{\rho}}{3}}}
\label{eq:frid2_m}
\end{equation}
The quantity $\rho$ is the mean density of the observed Universe whereas the terms with $\Lambda(t,r)$
and $\ddot{b}$ depend on a varying gravitational mass of the BBH located near the observed horizon. 
Using equation~(\ref{eq:lambda_f1}) we get $\ddot{b}=
\Lambda(t,r)c^{2}$. Then equation~(\ref{eq:frid2_m}) can be rewritten as
\begin{equation}
\frac{\ddot{a}}{a}=-\frac{\Lambda(t,r)c^{2}}{6}+\frac{8{\pi}G{\rho}}{3}+
\frac{\frac{{\dot\Lambda(t,r)}c^{2}}{6}+\frac{4{\pi}G{\dot{\rho}}}{3}}{\sqrt{\frac{\Lambda(t,r)c^{2}}{3}+\frac{8{\pi}G{\rho}}{3}}}
\label{eq:frid2_m2}
\end{equation}

The second modified Friedmann equation is quite similar to the classical equation~(\ref{eq:frid2}), where 
the cosmological function $\Lambda(t,r)$ is given by (\ref{eq:lambda_f}).

Let us consider the case when the term with the cosmological function dominates the term with mean density
of the Universe. Assuming an exponential change of the mass of the BBH (\ref{eq:force}), we get 
$\dot{\Lambda}(t,r)=\alpha \Lambda(t,r)$. Then, equation~(\ref{eq:frid2_m2})
transforms into the following equation:
\begin{equation}
\frac{\ddot{a}}{a}=-\frac{\Lambda(t,r)c^{2}}{6}-\frac{\alpha}{2}\sqrt{\frac{\Lambda(t,r)c^{2}}{3}}
\label{eq:frid2_mm}
\end{equation}

Using equation~(\ref{eq:frid1_m}), we get $\Lambda(t,r)c^{2}=3(H^{2}+H\dot{b})$, where $H=\frac{\dot{a}}{a}$
is the Hubble constant. Then, equation~(\ref{eq:frid2_mm}) can be rewritten as follows:
\begin{equation}
\frac{\ddot{a}}{a}=-\frac{1}{2}(H^{2}+H\dot{b}+{\alpha}\sqrt{H^{2}+H\dot{b}})\approx -\frac{1}{2}(H^{2}+{\alpha}H)
\label{eq:frid2_mH}
\end{equation}
When we write the approximate expression we assume that we can neglect the term $H\dot{b}$ as compared with others. 
It follows from equation~(\ref{eq:frid2_mH}) that acceleration for a comoving observer is positive 
$\ddot{a}>0$ if $\alpha < -H$, that there is sufficiently fast increase of the gravitational mass of the BBH.
It follows from equation~(\ref{eq:lambda_f1}) that the second term in the second Friedmann 
equation~(\ref{eq:frid2_mm}) depends on the distance to the BBH as a square root
$\sqrt{\Lambda(t,r)c^{2}/3} \propto \sqrt{r_{0}/r}$. The weaker dependence on the distance
as compared with the first term results in prevailing the second term on large distances over the first term
thus determining the dynamics of expanding of the Universe.

\section{Discussion}

Equation~(\ref{eq:lambda_f2}) shows that the cosmological function is expressed through a dimensionless
quantity $f^{2}r_{0}/r$ composed of two parameters of our model $f$ and ${r_{0}}/{r}$. 
The cosmological function is equal to 
the observed value of the cosmological constant (\ref{eq:lambda}) if 
\begin{equation}
f^{2} \frac{r_{0}}{r} \approx 1.6
\label{eq:param}
\end{equation}
Our basic assumption of weak gravitational fields means that $r_{0}/r \ll 1$, i.e. an observer 
is located far away from the BBH. Then $f \gg \sqrt{1.6} \approx 1.3$. Assuming that a value of 
$f\sim10$, we get from (\ref{eq:param}) the following estimate ${r_{0}}/{r}\sim0.02$.
If we assume that the distance, $r \sim 50$ billion light-years, i.e. the size of the Universe
 by the order of the magnitude \citep{Davis}, we get the following estimate of the radius of the BBH: 
$r_{0} \sim1$ billion light-years. This radius corresponds to the mass 
of the BBH of $\sim 6\times 10^{54}$ gm.               

Using equation ~(\ref{eq:param}) we can now validate our assumption of negligibly small
derivatives of $a(t,r)$ over space coordinates as compared with those of $b(t,r)$, that is
\begin{equation}
b^{\prime\prime} \gg a^{\prime\prime} \ and\ b^{\prime\prime} \gg b^{\prime} a^{\prime} 
\label{eq:estim1}
\end{equation}
where the prime denotes a derivative over a space coordinate.
The relationships of equation ~\ref{eq:estim1} can be estimated as follows:
\begin{equation}
(\frac{\alpha}{c})^2 \frac{r_0}{r}\gg \frac{1}{r^2}\ and\ (\frac{\alpha}{c})^2 \frac{r_0}{r}\gg \frac{\alpha}{c} \frac{r_0}{r}\frac{1}{r}
\label{eq:estim2}
\end{equation}
Substituting equation ~(\ref{eq:param}) into equation ~\ref{eq:estim2}, we obtain the following estimates:
\begin{equation}
\frac{1.6}{(cT)^2}\gg \frac{1}{r^2} \ and\ \frac{f}{cT} \gg \frac{1}{r}\ or\ {\alpha} \gg \frac{c}{r}
\label{eq:estim3}
\end{equation}
The conditions of equation ~(\ref{eq:estim3}) are satisfied for $r \sim 50$ billion light-years (a size of the Universe) and $f=10$.

There is an idea that a fraction of the black holes survives the stage of maximal contracting of the universe \citep{Clifton}. These relict black holes can produce the supermassive black holes in the galaxy centres and are responsible for the effects of 'dark matter'. This idea is in a qualitative agreement with our assumption about the existence of the BBH.

The first Friedmann equation (\ref{eq:frid1}) characterizes an isotropic and homogeneous model
of the Universe. All locations of an observer are equivalent. It can be shown \citep{McVittie2}
that galaxies move away from each other similarly in all comoving coordinate systems with
the same expansion rate defined by the Hubble constant. The modified Friedman equations 
with the cosmological function (\ref{eq:lambda_f}) are derived for two combined metrics: the basic FRW metric
and the additional modified Schwarzschild metric. These metrics are considered in a single system of coordinates. 
If an observer moves to another coordinate system, the traditional terms of the Friedmann equations
should not change; they are derived for the classical isotropic and homogeneous universe. 
However, the cosmological function $\Lambda(t,r)$ should change because it depends on a distance, $r$,
from the origin of the Schwarzschild system of coordinates. The Schwarzschild metric is a kind of 
the anisotropic background for the shifted observer.
Comparing the modified Friedmann equations with the classical ones, we conclude that the modified 
equations are valid for an observer at any location. However, the new cosmological term that depends on the distance from 
the BBH thus changes when the observer moves from one location to another. Thus, positive relative acceleration of 
observable galaxies is result of difference in deceleration of galaxies in nonstationary gravitational field of the BBH.

This paper is devoted to a mathematical solution of the Einstein equations for an expanding universe with a varying gravitational mass. This solution allows us to generalize the Friedmann equations for the varying gravitational mass. A consideration of observational cosmological aspects (CMB spectra, generation of large structures of the Universe, etc.) is beyond of the scope of this work. 

\section{Conclusions}

We generalized the Friedmann equations for a model of the Universe with a varying
gravitational mass. We managed to derive an expression for the cosmological constant directly from the Einstein equations.
The model leads to the following conclusions:

(i) The modified Friedmann equation is equivalent to the classical Friedmann equations when the perturbation $b(r,t)\rightarrow 0$. This allows us to suggest that the modified Friedmann equations do not contradict to all observational cosmological effects. A major distinction of the modified Friedmann equations from the classical equations is that the modified equations give us an expression for the cosmological constant based on physical quantities that is derived from the Einstein equations. 
Estimates of the cosmological constant based on this expression give values that are quantitatively close to the observed value. 

(ii) The cosmological principle of isotropy and homogeneity of the Universe
is not universal. The Universe is isotropic and homogeneous locally only.
Possible effects of anisotropy and inhomogeneity of the Universe can be observable for distant objects.

(iii) A disturbance of the FRW metric is caused by a very big black hole (BBH) located near 
the observed horizon of the Universe. The disturbance of the FRW metric results
in an effective cosmological constant in the modified Friedmann equations.
This effective cosmological constant is analogous to the dark energy term in the
classical Friedmann equations. Estimates of this effective cosmological constant
show that its value is close to observed cosmological constant by the order of magnitude
provided reasonable assumptions about the dimensionless parameters of the model have been made.

(iv) The observed positive acceleration of the expansion of the Universe corresponds to
an increase of the gravitational mass of the BBH. In the future, this increase of the BBH can 
stop the expansion of the Universe and cause its collapse. We believe that a cyclic model of Universe can 
be developed on the basis of cyclic transformation of black holes, baryons, and electromagnetic and gravitation radiation.

Our conclusions are supported by a few observations that are related to anisotropic and inhomogeneous
effects in cosmology (
\citep{Stavrinos, Erickcek, Hoftuft, Ade2, Schwarz, Javanmardi, Zhao, Colin, Riess2}). 
Besides that, a change of the gravitational mass of the Universe could be verified by
observations of the monopole component of non-stationary gravitational field.
What would be the waveform of the monopole component,
accompanying the detected transverse gravitational waves \citep{Abbot}? 
The LIGO design is specifically aimed at the transverse waves, but perhaps the NANOGrav
approach with the pulsar timing \citep{Arzoumanian} might be sensitive to the monopole signal?

\section*{Acknowledgements}

The authors thank John Mather, Sergei Kopeikin, Alexey Bogomazov, Dmitry Makarov and
 anonymous reviewer for helpful discussions.






\appendix

\section{Derivation of the modified Friedmann equations}

Let us consider the Einstein equations~(\ref{eq:einst}) without the cosmological
constant and derive the Friedmann equations for the disturbed FRW metric (\ref{eq:schwarz2})
with a function $b(t,r)$ assumed to be small. Thus, we neglect all
the products of the Christoffel symbols that are proportional to the squared function $b(t,r)$.
Elsewhere we also neglect all terms that contain a non-linear combination of $b(t,r)$.
The zero component of the Einstein equations is
\begin{equation}
R_{00}-\frac{1}{2}g_{00}R=-\frac{8 \pi G}{c^{4}}T_{00}
\label{eq:einst0_a}
\end{equation}
We can get the following expression for
the zero-index component of the left-hand side of the Einstein equations~(\ref{eq:einst})
(see chapter 100 in \citet{Tolman2}) :
\begin{equation}
\begin{split}
&R_{00}-\frac{1}{2}g_{00}R=-(\frac{1}{2g_{11}}\frac{\partial g_{11}}{c\partial t})
(\frac{1}{2g_{22}}\frac{\partial g_{22}}{c\partial t})-\\
&-(\frac{1}{2g_{11}}\frac{\partial g_{11}}{c\partial t})
(\frac{1}{2g_{33}}\frac{\partial g_{33}}{c\partial t})-
(\frac{1}{2g_{22}}\frac{\partial g_{22}}{c\partial t})
(\frac{1}{2g_{33}}\frac{\partial g_{33}}{c\partial t})-\\
&-g_{00}(\frac{1}{2g_{11}g_{22}}\frac{\partial^2 g_{22}}{\partial x_{\ast}^2}
+\frac{1}{2g_{11}g_{33}}\frac{\partial^2 g_{33}}{\partial x_{\ast}^2}+
\frac{1}{2g_{11}g_{22}}\frac{\partial^2 g_{11}}{\partial y_{\ast}^2}+\\
&+\frac{1}{2g_{22}g_{33}}\frac{\partial^2 g_{33}}{\partial y_{\ast}^2}+
\frac{1}{2g_{11}g_{33}}\frac{\partial^2 g_{11}}{\partial z_{\ast}^2}+
\frac{1}{2g_{22}g_{33}}\frac{\partial^2 g_{22}}{\partial z_{\ast}^2})
\label{eq:R00_a}
\end{split}
\end{equation}
A combination of the second derivatives in~(\ref{eq:R00_a}) is an effective cosmological
constant, which can be named as the 'cosmological function' because it depends on time and 
spatial coordinate:
\begin{equation}
\begin{split}
&\Lambda(t,r)=\frac{1}{2g_{11}g_{22}}\frac{\partial^2 g_{22}}{\partial x_{\ast}^2}
+\frac{1}{2g_{11}g_{33}}\frac{\partial^2 g_{33}}{\partial x_{\ast}^2}+
\frac{1}{2g_{11}g_{22}}\frac{\partial^2 g_{11}}{\partial y_{\ast}^2}+\\
&+\frac{1}{2g_{22}g_{33}}\frac{\partial^2 g_{33}}{\partial y_{\ast}^2}+
\frac{1}{2g_{11}g_{33}}\frac{\partial^2 g_{11}}{\partial z_{\ast}^2}+
\frac{1}{2g_{22}g_{33}}\frac{\partial^2 g_{22}}{\partial z_{\ast}^2}
\label{eq:lambda_a}
\end{split}
\end{equation}
It should be noted that $g_{00} \approx -1$ in our approximation. Equalling (\ref{eq:R00_a}) to
the zero-index component of the energy-momentum tensor, we get the following expression:
\begin{equation}
3[(\frac{\dot{a}}{a})^{2}+\frac{\dot{a}}{a}\dot{b}]={\Lambda(t,r)}c^2+8{\pi}G{\rho}
\label{eq:fr1m_a}
\end{equation}
The terms on the left-hand side of equation~(\ref{eq:fr1m_a}) are derived from the products
of the derivatives over time in~(\ref{eq:R00_a}). The additional term with $\dot{b}$
is significant at differentiating equation~(\ref{eq:fr1m_a}) over time to derive
the second Friedman equation. 
Solving equation~(\ref{eq:fr1m_a}) with
respect to the ratio, $\frac{\dot{a}}{a}$, we get the following expression:
\begin{equation}
\frac{\dot{a}}{a}=\pm\sqrt{\frac{\Lambda(t,r)c^{2}}{3}+\frac{8{\pi}G{\rho}}{3}+\frac{\dot b^2}{4}}-\frac{\dot b}{2}
\label{eq:dota_a}
\end{equation}
The upper sign in equation~(\ref{eq:dota_a}) and elsewhere corresponds to the case of an expanding universe
(the Hubble constant is positive) and the bottom sign corresponds to a collapsing universe
(the Hubble constant is negative). Squaring~(\ref{eq:dota_a}) we get 
the first modified Friedmann equation:
\begin{equation}
(\frac{\dot{a}}{a})^{2}=\frac{\Lambda(t,r)c^{2}}{3}+\frac{8{\pi}G{\rho}}{3}+\frac{\dot b^2}{2} \mp \dot b \sqrt{\frac{\Lambda(t,r)c^{2}}{3}+\frac{8{\pi}G{\rho}}{3}+\frac{\dot b^2}{4}}
\label{eq:frid1m_a}
\end{equation}
To get the second modified Friedmann equation, let us differentiate equation~(\ref{eq:dota_a})
with respect to time. Then using equation~(\ref{eq:frid1m_a}) we get 
the second modified Friedmann equation:
\begin{equation}
\begin{split}
&\frac{\ddot{a}}{a}=\frac{\Lambda c^{2}}{3}+\frac{8{\pi}G{\rho}}{3}+\frac{\dot b^2-\ddot b}{2}\mp
\frac{\frac{c^2}{3}(\dot b \Lambda-\frac{\dot \Lambda}{2})+\frac{8\pi G}{3}(\dot b \rho-\frac{\dot \rho}{2})+\frac{\dot b}{4}(\dot b^2-\ddot b)}
{\sqrt{\frac{\Lambda c^{2}}{3}+\frac{8{\pi}G{\rho}}{3}+\frac{\dot b^2}{4}}} 
\label{eq:frid2m_a}
\end{split}
\end{equation}

For weak gravitational fields $b(t,r) \ll 1$, we can neglect all terms with
 $\dot{b}$ and $\ddot{b}$ except for the term $-\ddot{b}/2$. In this approximation 
we get the second modified Friedmann equation in the form of equation~(\ref{eq:frid2_m})
that is provided in the main text.


\bsp	
\label{lastpage}
\end{document}